\documentclass[fleqn,usenatbib]{mnras}
\usepackage{newtxtext,newtxmath}
\usepackage{mathptmx}
\usepackage{txfonts}
\usepackage[T1]{fontenc}

\DeclareRobustCommand{\VAN}[3]{#2}
\let\VANthebibliography\thebibliography
\def\thebibliography{\DeclareRobustCommand{\VAN}[3]{##3}\VANthebibliography}
\usepackage{graphicx}	
\usepackage{amsmath}	
\usepackage{multicol} 
\usepackage{pdflscape}

\title[Chemical anomaly search]{A Search for Chemical Anomalies of Seven A-types Stars}

\author[ Y. Nasolo et al.]{
Yahya Nasolo,$^{1}$\thanks{E-mail: nasolo@ankara.edu.tr}
and Şeyma Çalışkan$^{1}$
\\
$^{1}$Department of Astronomy and Space Sciences, Ankara University, Tandoğan, Ankara 06100, Türkiye
}
\date{Accepted XXX. Received YYY; in original form ZZZ}
\pubyear{2022}

\begin{document}
\label{firstpage}
\pagerange{\pageref{firstpage}--\pageref{lastpage}}
\maketitle

\begin{abstract}
We present a chemical abundance analysis of seven A-type stars with no detailed chemical abundance measurements in the literature. High-resolution spectra of the targets – HD 2924, HD 4321, HD 26553, HD 125658, HD 137928, HD 154713, and HD 159834 – were obtained using the Coudé Echelle Spectrograph at the TÜBİTAK National Observatory. We determined the atmospheric abundances of the samples and measured the elemental abundances of C, N, O, Na, Mg, Al, Si, S, K, Ca, Sc, Ti, V, Cr, Mn, Fe, Co, Ni, Cu, Zn, Sr, Y, Zr, Ba, La, Ce, Nd, Sm, Eu, and Gd. The masses of the stars were estimated based on their evolutionary tracks, and their ages were calculated using isochrones. We also calculated the radii of the stars. The abundance patterns of HD 4321, HD 125658, and HD 154713 were found to be in agreement with those of classical Am stars, with underabundant Ca and Sc, overabundant heavier elements, and moderate overabundance of iron-peak elements. We found that HD 137928 and HD 159834 have similar abundance characteristics to marginal Am-type stars. The elemental distributions of HD 2924 and HD 26553 are consistent with the pattern of normal A-type stars. The iron, nickel, and zinc abundances of HD 125658 and HD 137928 are significantly higher than other Am stars. These values suggest that they are among the most metal-rich Am stars.  
\end{abstract}

\begin{keywords}
Stars: chemically peculiar – abundances -- HD~2924 -- HD~4321 -- HD~26553 -- HD~125658 -- HD~137928 -- HD~154723 -- HD~159834
\end{keywords}



\section{Introduction}
\label{sec:intro}
In the course of stellar evolution, stars inherit the chemical composition of their birthplace; however, the atmospheric chemical abundance distribution changes over time due to poorly understood internal stellar processes. The warmer post-ZAMS mid-massive stars, corresponding to mid-B through early-F spectral type stars, attract particular attention since a substantial fraction exhibit chemical abundance anomalies, which reflect these internal processes. Detailed chemical abundance analysis of these stars is essential since it potentially provides a basis for understanding such phenomena. The chemically peculiar (CP) metallic-line (Am) stars are the typical targets for these studies.
CP stars are characterized by enhanced absorption lines \citep{Preston1974, Monier2019}. Am stars are A- to early F-type stars sharing the same region with the normal A-type stars, with effective temperature (${T}_\text{eff}$) between 7000 and 9000 \,K. They lie in the cooler portion of the main-sequence CP stars. The Am stars were first recognized as a group by \citet{Titus1940}. \citet{Preston1974} classified the CP stars into four groups, with Am stars being the CP1 group.
Am stars are characterized by three spectral characteristics: the strength of their Balmer lines; their  Ca~II~K lines, which are weaker than those of normal A-type stars with the same temperature; and their enhanced metallic lines relative to their normal counterparts.  In typical Am stars spectra, the intensity of the  Ca~II~K line is weaker than Balmer lines and resembles an earlier spectral type. The metallic lines are more intense and resemble a later spectral type \citep{Roman48}. \citet{Abt76}stated that the  Ca~II~K line in the spectra of such stars resembles the spectral type at least five subtypes earlier. The Am stars with  Ca~II~K and metallic lines differing by less than five subtypes are referred to as marginal Am stars \citep{trust2021}. Underabundance of Sc~II and/or Ca~II is a discovery criterion for this class \citep{Michaud2015}.
Based on observational statistics, the majority of Am stars are slow rotators \citep{Fossati2008}. This supports the hypothesis of chemical separation due to radiative diffusion and gravitational settling, since slow rotation is expected not to destroy the separation phenomenon \citep{Michaud2015}. Chemical separation requires calm atmospheres, which are expected given the slow rotation of these stars \citep{Takeda2008, Fossati2008,Abt2009}. In close binaries, slow rotation is associated with tidal synchronization \citep{Khokhlova1981}. However, observations of low-rotating normal A-type stars pose a challenge to this view of the impact of rotation. Normal A-type stars exhibit nearly solar abundances with the heavier elements falling within the range of $\pm{0.40}$\,dex relative to the Sun \citep{Adelman2007}. This paper focuses on determining in detail the chemical abundances and fundamental parameters (e.g. effective temperatures and surface gravities) of the sample stars. 
The paper is organized as follows: Section~\ref{sec:obs} describes the target selection, observations, and data reduction. We present the determination of atmospheric stellar parameters and the procedure used to perform the abundance analysis in Section~\ref{sec:atm_abu}. The evolutionary status of the stars is shown by employing the HR diagram in Section~\ref{sec:hr_dia}. We discuss the results in Section~\ref{sec:res}, and we conclude in Section~\ref{sec:conc}.

\section{OBSERVATION DATA}
\label{sec:obs}
\subsection{Target selection}
One of the dominant mixing processes that suppress the effect of chemical separation in atomic diffusion, is linked to rotational velocity. A statistical analysis of observations suggests that all main-sequence stars of the targeted temperature regime with equatorial rotational velocities less than $80$\,km\,s$^{-1}$ are Am \slash Fm stars \citep{Abt2000}. \citet{AbtMoyd1973} initially reported that the equatorial velocity of observed CP1 stars is less than $120$\,km\,s$^{-1}$. This observation is consistent with chemical separation due to atomic diffusion occurring when the He~II convection zone disappears at equatorial velocities lower than $70-120$\,km\,s$^{-1}$ \citep{Michaud1982}. We therefore selected targets in this temperature regime with low rotation velocities ($v$ sin $i$ < $70$\,km\,s$^{-1}$) and whose chemical abundance has not been previously derived.

\subsection{Observations and data reduction}
\begin{figure}
	\includegraphics[width=\columnwidth]{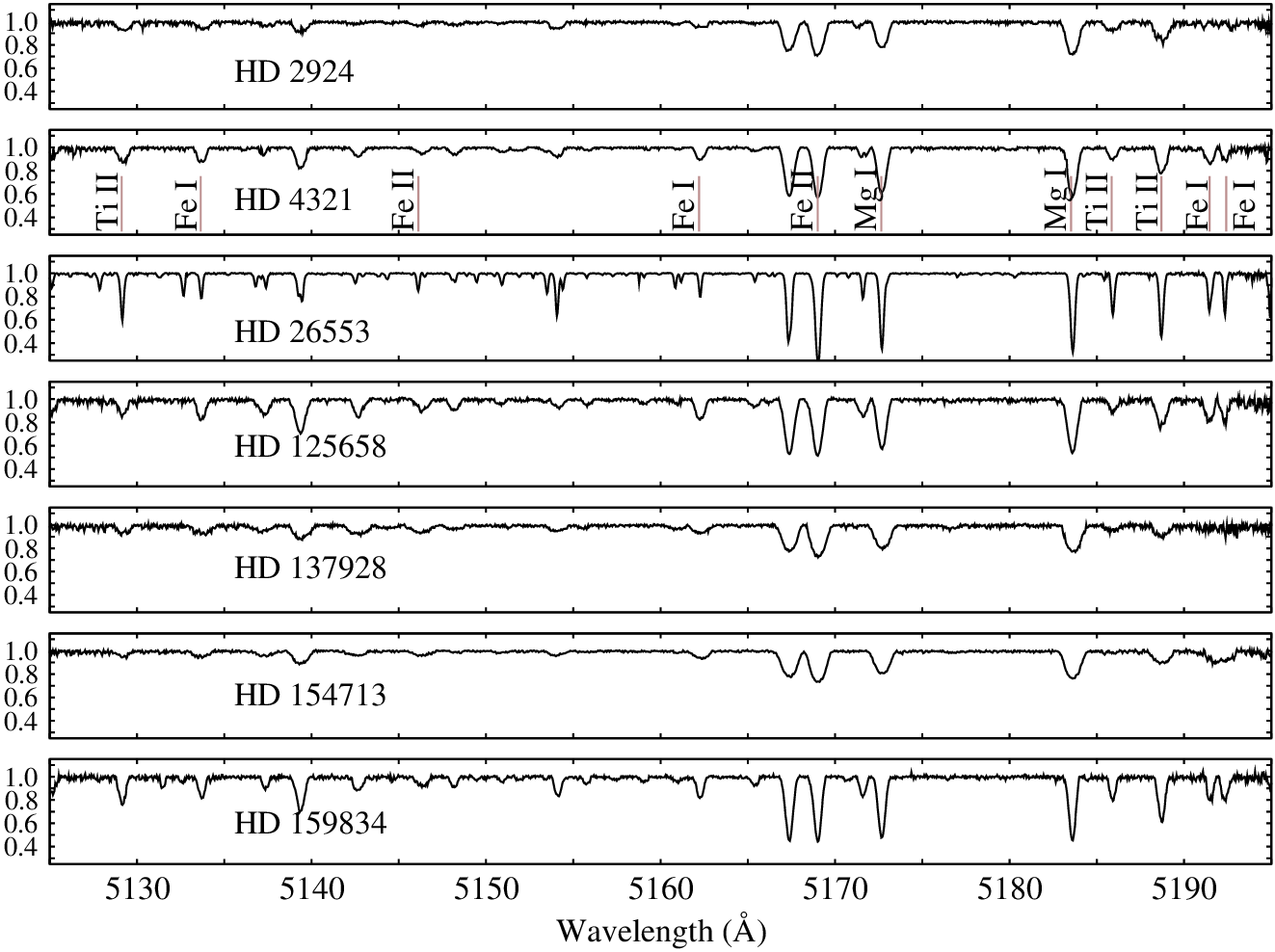}
    \caption{A sample of normalized orders of the target stars and some of the identified unblended lines used for chemical abundance determination.}
    \label{fig:spectra}
\end{figure}

For this study, we used the high-resolution R$\sim$40,000 spectroscopic data obtained by the Coudé Echelle Spectrograph attached to the 1.5~m Russian–Turkish Telescope (RTT-150) at TÜBİTAK National Observatory (TUG). The spectra cover a wavelength range of 3900 to 8770~\AA. Along with the target star’s spectra, each observation run began with acquiring calibration frames and arc spectra: the halogen lamp for flat fielding and the Th–Ar lamp for wavelength calibration.
Table~\ref{tab:obs_log} lists the observation log of our targets, including observation times, exposure times, numbers of co-added spectra, and signal-to-noise ratios (SNR). The raw spectra were reduced using IRAF’s standard reduction routines. Initially, image header information was methodized to allow processing in the IRAF environment. The reduction process started with bias subtraction followed by flat fielding. The next step was the removal of scattered light from star spectra. Using the Th-Ar spectra, we conducted wavelength calibration and lastly performed cosmic ray removal. We measured the Doppler shift on the spectra using unblended moderately strong absorption lines. Then, the heliocentric radial velocities (${V}_\text{helio}$) were determined and necessary Doppler corrections were applied to all spectra. 
For six of the targets, we acquired two spectral frames, which were combined to improve the SNR. We could not acquire a second frame for HD~125658 due to adverse weather. The echelle orders of the spectra were normalized according to spline-fitted continuum points. Figure~\ref{fig:spectra} shows a sample of normalized orders of the spectra of the stars at about 5100~\AA~, as well as some of the identified unblended lines used for chemical abundance determination. The lines clearly exhibit the broadening effect of rotational velocity.

\begin{table*}
 \caption{The observation log of the target stars. The V-band magnitude were estimated by \citet{Oja1991}}
 \label{tab:obs_log}
 \begin{tabular}{l c c c c }

		\hline
		Star name & V-mag & Exposure time & Observation time & SNR \\
		 & (mag) & (s) & (JD) & (@5000-7000~\AA)\\
		\hline
		HD~2924 & 6.700 [0.009] & 4500 x 2 & 2457676.37750 & 297\\
		HD~4321 & 6.510 [0.009] & 4500 x 2 & 2457675.50366 & 250\\
		HD~26553 & 6.080 [0.009] & 4500 x 2 & 2455554.33347 & 287\\
		HD~125658 & 6.445 [0.010] & 4500 x 1 & 2458232.38508 & 115\\
		HD~137928 & 6.430 [0.010] & 4500 x 2 & 2458233.33802 & 280\\
		HD~154713 & 6.441 [0.010] & 4500 x 2 & 2458234.41497 & 302\\
		HD~159834 & 6.100 [0.010] & 4500 x 2 & 2458232.51027 & 362\\
		\hline
 \end{tabular}
\end{table*}

\section{ATMOSPHERE PARAMETERS AND CHEMICAL ABUNDANCES}
\label{sec:atm_abu}
Spectroscopy and photometry are the traditional methods used to determine stellar atmospheric parameters. In this work, we employed both methods. We used the photometric method to find the initial atmospheric parameters. Then, we refined the parameters with the spectroscopic method.

\subsection{Line identification and equivalent width measurement}
We used line lists from the Kurucz line database \citep{Kurucz1995} and the Vienna Atomic Line Database (\citet{Piskunov1995}; \citet{kupka1999}; \citet{Ryabchikova1999}). Line identification and equivalent width measurements of the atomic species found in the spectra of the target stars were implemented through an IDL visualization program, BINMAG6 \citep{Kochukhov2018}. Equivalent width measurements were performed by Gaussian profile fitting of the lines. Table~\ref{tab:line_list} lists the measured spectral lines.

\begin{table*}
 \caption{ List of spectral lines used in the abundance analysis of target stars. The full table is available online. A portion of the table is shown here}
 \label{tab:line_list}
 \begin{tabular}{l c c c c c c c c c}
 \hline
Atomic Sp. & Wavelength & log $gf$  & Equivalent width (pm)                                                                  \\
           &   (nm)     &        & HD2924  & HD4321  & HD26553 & HD125658 & HD137928 & HD154713 & HD159834 \\
 \hline
   C I     & 477.0027   & -2.722 & $\dots$ & $\dots$ & $\dots$ & $\dots$  & $\dots$  & $\dots$  & 01.50    \\
   C I     & 477.1742   & -2.120 & $\dots$ & 06.46   & $\dots$ & 05.87    & 02.93    & 03.78    & 04.22    \\
   C I     & 477.5897   & -2.163 & 02.98   & 05.38   & 01.60   & 03.09    & $\dots$  & $\dots$  & $\dots$  \\
   C I     & 493.2049   & -1.884 & 04.45   & 06.89   & 03.04   & 03.56    & $\dots$  & 02.74    & 02.90    \\
   C I     & 502.3849   & -2.400 & $\dots$ & 02.25   & 00.71   & $\dots$  & $\dots$  & $\dots$  & 02.02    \\
  \hline
 \end{tabular}
\end{table*}
\subsection{Stellar parameters}
Photometric stellar parameters were determined using the Strömgren $ubvy\beta$ photometric indices of \citet{Paunzen2015}. The photometric data were used in the calibrations performed by \citet{Napiwotzki1993} to determine the ${T}_\text{eff}$ and surface gravity (log $g$) of each target. These calibrations have systematic and statistical errors of 200\,K for ${T}_\text{eff}$ and 0.10\,dex for log $g$ \citep{Napiwotzki1993}. The computed stellar parameters are given in Table~\ref{tab:photometrically}. 

\begin{table}
	\centering
	\caption{Photometrically determined stellar parameters}
	\label{tab:photometrically}
	\begin{tabular}{lccc} 
		\hline
		Star name & ${T}_\text{eff}$ & log $g$ \\
	     & [K] & [dex] \\
		\hline
	    HD~2924 & 8633 & 3.45  \\
		HD~4321 & 7659 & 3.27  \\
		HD~26553 & 7783 & 2.41 \\
		HD~125658 & 7966 & 3.94 \\
		HD~137928 & 8382 & 3.81 \\
		HD~154713 & 8217 & 3.63 \\
		HD~159834 & 7799 & 3.57 \\
		\hline
	\end{tabular}
	\\
\end{table}

\textit{Spectroscopic atmospheric parameters} determination. One of the classic methods used to yield the effective temperature spectroscopically is to employ Fe~I lines throughout the stellar spectrum and iterate the effective temperature until their abundances show no trend with the excitation potential (EP) of each line. In contrast with photometric methods, no reddening information is needed for this technique; however, high-resolution spectra with a resolving power of R$\sim \lambda/\Delta \lambda$ $\geq$ 15,000 are required, preferably with wide wavelength coverage throughout the optical range \citep{Frebel2013}.
The different methods are known to produce temperatures with systematic offsets from each other. We used the photometrically determined atmospheric parameters ((${T}_\text{eff}$, log $g$) to construct initial model atmospheres used in the EP method, where ${T}_\text{eff}$ was derived from the excitation equilibrium of the Fe~I lines. The models were constructed using ATLAS9~\citep{Kurucz1993}. Depending on the ${T}_\text{eff}$ derived from the EP method, we derived the stellar surface gravities through Fe~I-II ionization equilibrium. The log~$g$ was determined from the balance of abundances derived from individual Fe~I-II while the value of log~$g$ was iterated. The microturbulent velocity ($\xi$) was calculated by adopting the values which produced no trend of equivalent widths with abundances derived from individual Fe~I lines. This allowed the chemical abundance of iron to be derived. We measured the $v$ sin $i$ of the target stars by comparing the observed spectra with the synthetic spectra. The synthetic spectra were produced using SYNTHE \citep{Kurucz2005,SKurucz1993}. All atmospheric parameters derived from spectroscopy are listed in Table~\ref{tab:spectroscopy}.

\begin{table*}
	\centering
	\caption{Atmospheric parameters from spectroscopic analysis}
	\label{tab:spectroscopy}
	\begin{tabular*}{\textwidth}{@{}l@{\hspace*{40pt}}c@{\hspace*{40pt}}c@{\hspace*{40pt}}c@{\hspace*{40pt}}c@{\hspace*{40pt}}c@{\hspace*{40pt}}c@{}}
		\hline
		Star name & ${T}_\text{eff}$ &  log $g$ & [Fe/H] & $\xi$ & ${V}_\text{helio}$ & $v$ sin $i$  \\
		 & [K] & [dex] & [dex] & [\,km\,s$^{-1}$] & [\,km\,s$^{-1}$]  & [\,km\,s$^{-1}$] \\
		\hline
		HD~2924 & 8650$\pm$150 & 3.70$\pm$0.10 & 0.12$\pm$0.18 & 2.60$\pm$0.5 & 01.90$\pm$0.25 & 30$\pm$5\\
		HD~4321 & 8500$\pm$150 & 4.00$\pm$0.10 & 0.14$\pm$0.15 & 3.00$\pm$0.5 & -07.60$\pm$0.70 & 24$\pm$4\\
		HD~26553 & 8350$\pm$150 & 2.20$\pm$0.10 & 0.08$\pm$0.19 & 1.90$\pm$0.5 &  -22.00$\pm$2.60 & 10$\pm$2\\
		HD~125658 & 8800$\pm$150 & 4.60$\pm$0.10 & 0.41$\pm$0.10 & 5.00$\pm$0.5 &  01.70$\pm$0.50 & 26$\pm$5\\
		HD~137928 & 9100$\pm$150 & 3.80±0.10 & 0.42$\pm$0.12 & 4.00$\pm$0.5 & -03.90$\pm$0.40 & 34$\pm$5\\
		HD~154713 & 8350$\pm$150 & 3.50$\pm$0.10 & 0.18$\pm$0.15 & 3.00$\pm$0.5 & -06.30$\pm$0.50 & 39$\pm$5\\
		HD~159834 & 8350$\pm$150 & 3.90$\pm$0.10 & 0.13$\pm$0.20 & 3.20$\pm$0.5 & -16.20$\pm$2.00 & 14$\pm$3\\
		\hline
	\end{tabular*}
\end{table*}

\subsection{Abundance analysis}
We derived the abundance of each atomic species using the measured equivalent widths in WIDTH9 \citep{WKurucz2005, Sbordone2004}, which uses ATLAS9 model atmospheres to compute a line formation in Local Thermodynamic Equilibrium. We avoided using lines of equivalent widths greater than 200~m\AA~ for the abundance analysis to avoid the use of saturated lines that bias the abundances. We included hyperfine structure (HFS) splitting for the V, Sc, Mn, Y, and Ba spectral lines. HFS split lines due to electromagnetic multipole interaction between the nucleus and electrons, which arises from the energy of the nuclear magnetic dipole moment interacting with the magnetic field generated by the electrons, and shifts the lines of different isotopes. HFS has a small broadening effect, shifting the lines and altering their shape, which may lead to erroneous interpretations of spectra with the equivalent width method.

\begin{figure}
	\includegraphics[width=\columnwidth]{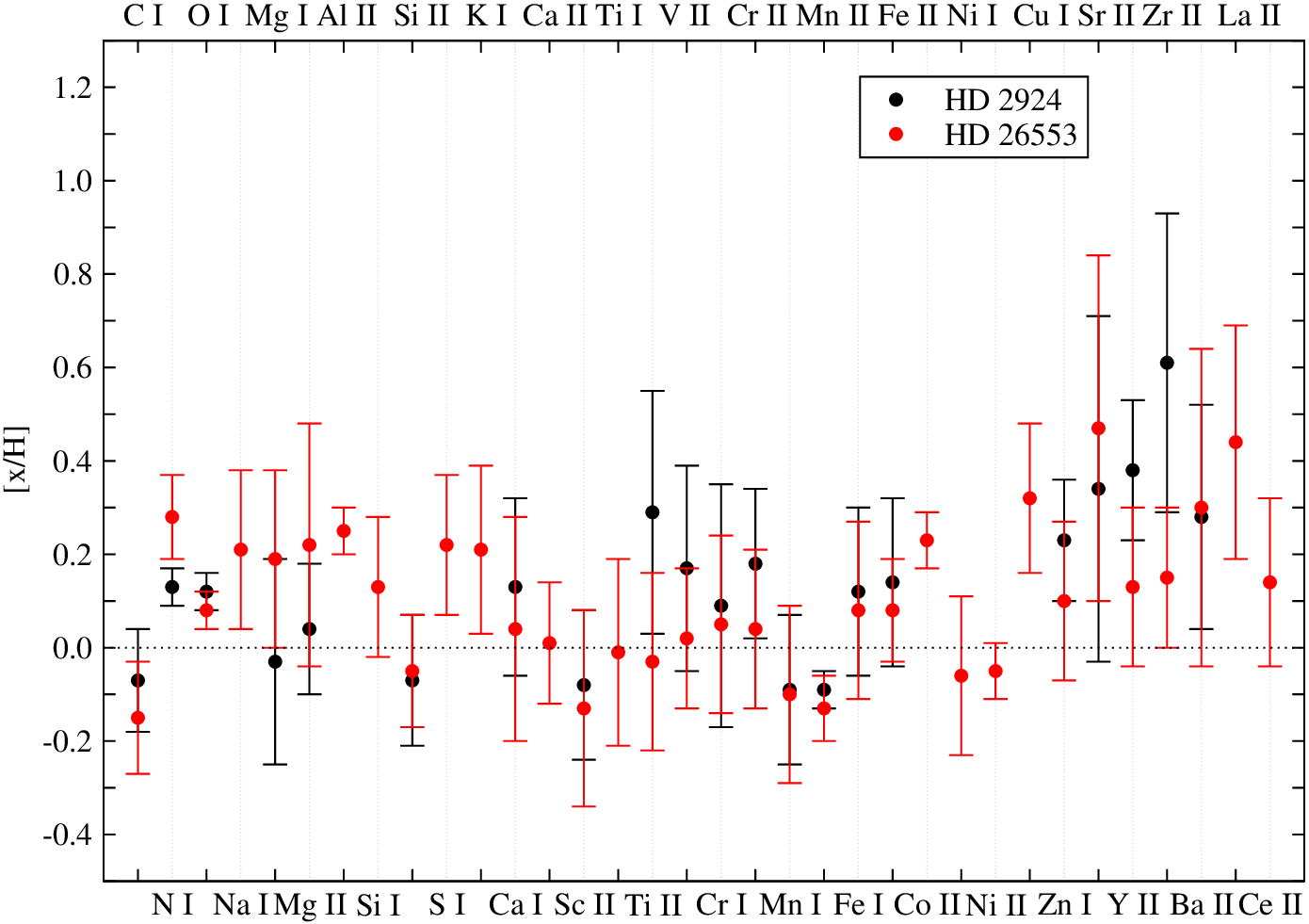}
    \caption{The differences between our derived chemical abundances and the solar abundances found by \citet{Grevesse1998} for HD~2924 and HD~26553}
    \label{fig:normal}
\end{figure}

\begin{figure}
	\includegraphics[width=\columnwidth]{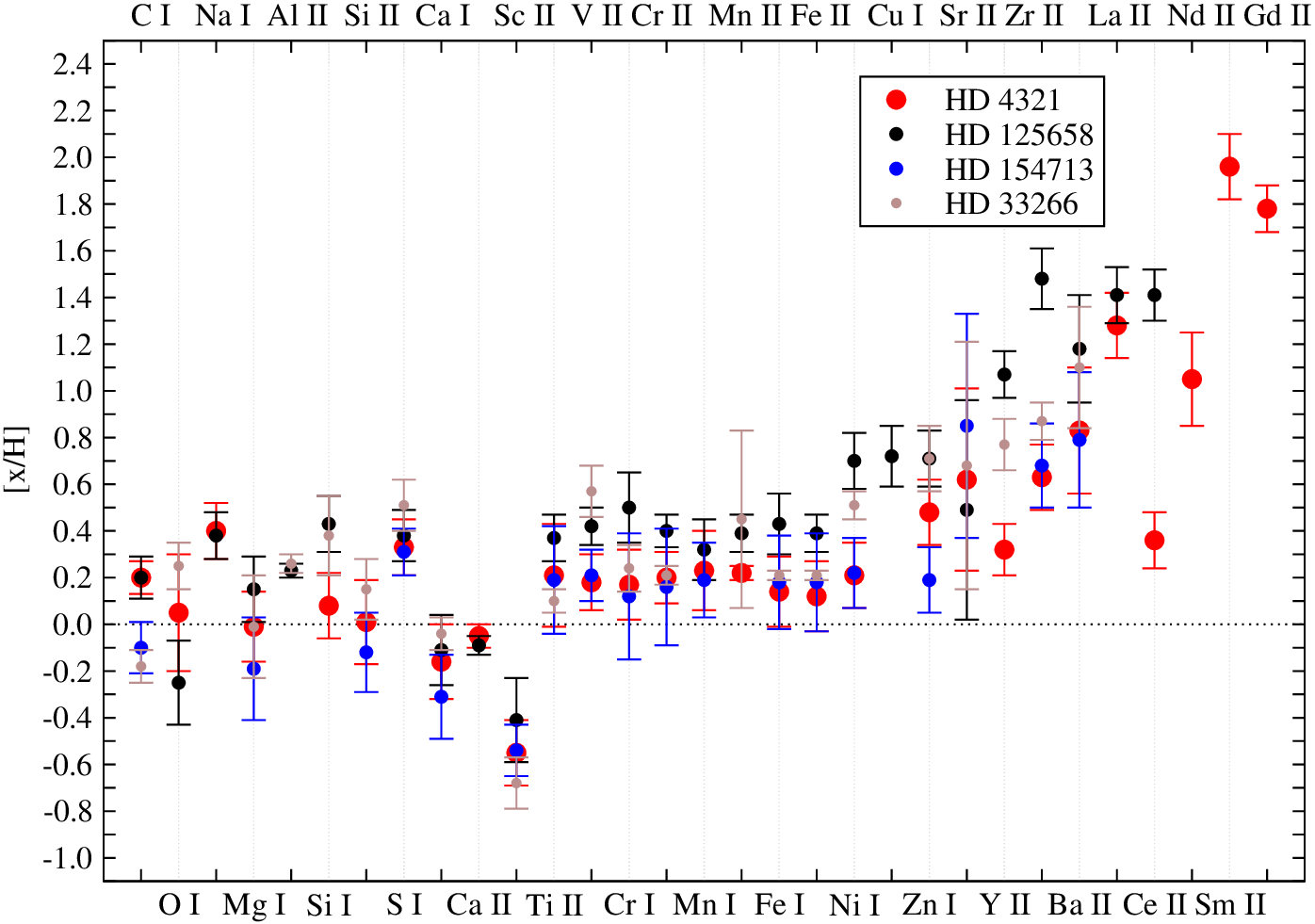}
    \caption{The differences between our derived chemical abundances and the solar abundances found by \citet{Grevesse1998} for the classic Am stars HD~4321, HD~125658, and HD~154713. HD~33266 is a classical Am star studied by \citet{caliskan2015}.}
    \label{fig:classic}
\end{figure}

\begin{figure}
	\includegraphics[width=\columnwidth]{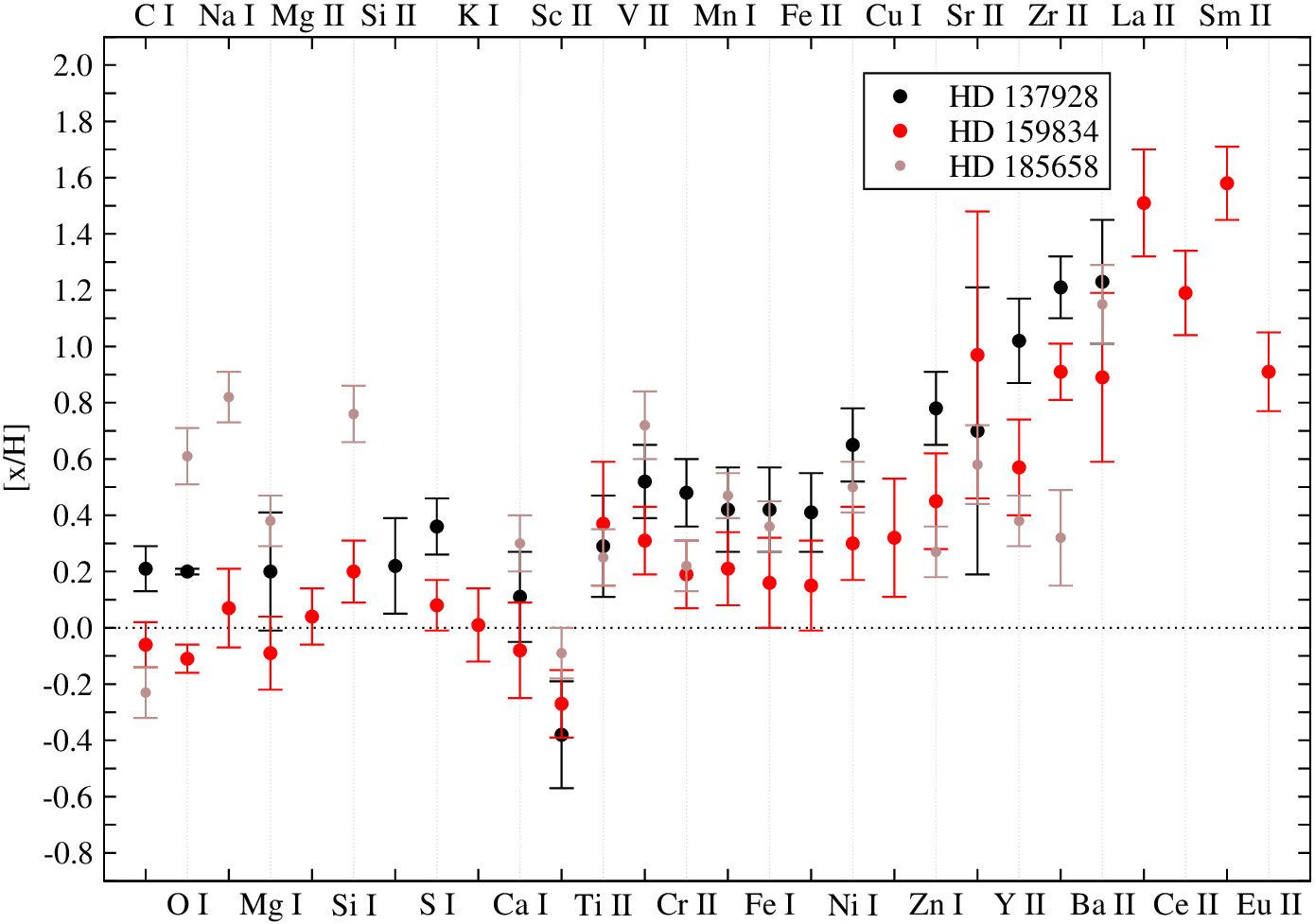}
    \caption{The differences between our derived chemical abundances and the solar abundances found by \citet{Grevesse1998} for the marginal Am stars HD~137928 and HD~159834. HD~185658 is a marginal Am star studied by \citet{trust2021}}
    \label{fig:marginal}
\end{figure}

To measure the abundance of HFS elements, we generated synthetic spectra using SYNTHE \citep{SKurucz1993, Kurucz2005} with HFS line lists from the Kurucz database, and we fitted them with the observed spectra lines. The spectra generation includes the broadening effects due to the instrumental profile and turbulent velocity. The abundance of the best fit between the observed and synthetic spectrum was adopted. The elemental abundances have random ($\sigma_r$) and systematic errors. The random errors are from the standard deviations of individual measurements. The systematic error of the chemical abundances due to atmosphere parameters is
\begin{equation}
 [\sigma_{sys} \text{ = }\sqrt{\sigma_{Teff}^2 + \sigma_{logg}^2 + \sigma_{\xi}^2}]
\end{equation}
The total error is
\begin{equation}
 [\sigma_{tot} \text{ = } \sqrt[3]{\sigma_{sys}^2 + \sigma_{r}^2} \text{ \slash }  \sqrt{N}]
\end{equation}  where N is the number of lines used to derive the chemical abundances.
We present the elemental abundances along with the number of lines and their total errors in Tables 4 and 5. The solar abundance ((log $\epsilon_{\sun}$) ) values are taken from \citet{Grevesse1998}. Figs~\ref{fig:normal}, \ref{fig:classic}, and \ref{fig:marginal} show the differential to solar abundances with total error bars. The abundance patterns of HD~4321, HD~125658, and HD~154713 show an agreement with the abundance patterns of classical Am stars, with underabundant Ca and Sc, overabundant heavier elements, and moderate over-abundance of iron-peak elements. We compared these patterns with the classical Am HD~33266 studied by \citet{caliskan2015}. We find that HD~137928 and HD~159834 have similar abundance characteristics to marginal Am-type stars. We also compared their patterns with the marginal Am HD 185658 (KIC~9349245) from \citet{trust2021}. The chemical element distributions of HD~2924 and the giant HD~26553 are consistent with the pattern of normal A-type stars.

\begin{landscape}
 \begin{table}
  \caption{Derived elemental abundances with the total error estimates of HD~2924 and HD~26553.}
  \label{tab:normaabu}
  \begin{tabular}{cccccc}
    \hline
    Atomic Sp. & HD~2924 & HD~2924 & HD~26553 & HD~26553 & Solar \\
             & log $\epsilon$ & [x/H] & log $\epsilon$ & [x/H] & log $\epsilon_{\sun}$ \\
    \hline
    C~I      & 8.45 & -0.07$\pm$0.11 (6) & 8.37 & -0.15$\pm$0.12 (19) & 8.52  \\
    N~I      & 8.05 & 0.13$\pm$0.04  (1) & 8.20 &  0.28$\pm$0.09 (2)  & 7.92  \\
    O~I      & 8.95 & 0.12$\pm$0.04  (1) & 8.91 &  0.08$\pm$0.04 (8)  & 8.83  \\
    Na~I     & $\dots$  &   $\dots$  & 6.54 &  0.21$\pm$0.17 (3)  & 6.33  \\
    Mg~I     & 7.55 & -0.03$\pm$0.22 (4) & 7.77 &  0.19$\pm$0.19 (4)  & 7.58  \\
    Mg~II    & 7.62 & 0.04$\pm$0.14  (1) & 7.80 &  0.22$\pm$0.26 (2)  & 7.58  \\
    Al~II    & $\dots$  &   $\dots$  & 6.72 &  0.25$\pm$0.05 (1)  & 6.47  \\
    Si~I     & $\dots$  &   $\dots$  & 7.68 &  0.13$\pm$0.15 (8)  & 7.55  \\
    Si~II    & 7.48 & -0.07$\pm$0.14 (2) & 7.50 &  -0.05$\pm$0.12 (2) & 7.55  \\
    S~I      & $\dots$  &   $\dots$  & 7.55 &  0.22$\pm$0.15(6)   & 7.33  \\
    K~I       & $\dots$  &   $\dots$  & 5.33 &  0.21$\pm$0.18(1)   & 5.12  \\
    Ca~I     & 6.49 & 0.13$\pm$0.19  (7) & 6.40 & 0.04$\pm$0.24 (17)  & 6.36  \\
    Ca~II    & $\dots$  &   $\dots$  & 6.37 & 0.01$\pm$0.13 (4)   & 6.36  \\
    Sc~II$^{HFS}$ & 3.09 & -0.08$\pm$0.16 (3) & 3.04 & -0.13$\pm$0.21 (11) & 3.17  \\
    Ti~I     & $\dots$  &   $\dots$  & 5.01 & -0.01$\pm$0.20 (4)  & 5.02  \\
    Ti~II    & 5.31 & 0.29$\pm$0.26  (1) & 4.99 & -0.03$\pm$0.19 (38) & 5.02  \\  
    V~II$^{HFS}$  & 4.17 & 0.17$\pm$0.22  (1) & 4.02 & 0.02$\pm$0.15 (9)   & 4.00  \\
    Cr~I     & 5.76 & 0.09$\pm$0.26  (2) & 5.72 & 0.05$\pm$0.19(9)    & 5.67  \\
    Cr~II    & 5.85 & 0.18$\pm$0.16 (10) & 5.71 & 0.04$\pm$0.17(39)   & 5.67  \\
    Mn~I$^{HFS}$  & 5.30 & -0.09$\pm$0.16 (1) & 5.29 & -0.10$\pm$0.19(8)   & 5.39  \\
    Mn~II$^{HFS}$ & 5.30 & -0.09$\pm$0.04 (1) & 5.26 & -0.13$\pm$0.07(7)   & 5.39  \\
    Fe~I     & 7.62 & 0.12$\pm$0.18 (42) & 7.58 & 0.08$\pm$0.19(202)  & 7.50  \\
    Fe~II    & 7.64 & 0.14$\pm$0.18 (46) & 7.58 & 0.08$\pm$0.11(102)  & 7.50  \\
    Co~II    & $\dots$  &   $\dots$  & 5.15 & 0.23$\pm$0.06(1)    & 4.92  \\
    Ni~I     & $\dots$  &   $\dots$  & 6.19 & -0.06$\pm$0.17(20)  & 6.25  \\
    Ni~II    & $\dots$  &   $\dots$  & 6.20 & -0.05$\pm$0.06(3)   & 6.25  \\
    Cu~I     & $\dots$  &   $\dots$  & 4.53 & 0.32$\pm$0.16(1)    & 4.21  \\
    Zn~I     & 4.83 & 0.23$\pm$0.13  (1) & 4.70 & 0.10$\pm$0.17(2)    & 4.60  \\
    Sr~II    & 3.31 & 0.34$\pm$0.37  (2) & 3.44 & 0.47$\pm$0.37(3)    & 2.97  \\
    Y~II$^{HFS}$  & 2.62 & 0.38$\pm$0.15  (1) & 2.37 & 0.13$\pm$0.17(5)    & 2.24  \\
    Zr~II    & 3.21 & 0.61$\pm$0.32  (1) & 2.75 & 0.15$\pm$0.15(8)    & 2.60  \\
    Ba~II$^{HFS}$ & 2.41 & 0.28$\pm$0.24  (1) & 2.43 & 0.30$\pm$0.34(4)    & 2.13  \\
    La~II    & $\dots$  &   $\dots$  & 1.61 & 0.44$\pm$025(2)     & 1.17  \\
    Ce~II    & $\dots$  &   $\dots$  & 1.72 & 0.14$\pm$0.18(1)    & 1.58  \\
    \hline
  \end{tabular}
 \end{table}
\end{landscape}

\begin{landscape}
 \begin{table}
  \caption{Derived elemental abundances with the total error estimates of HD~4321, HD~125658, HD~137928, HD~154713 and HD~159834.}
  \label{tab:cpabu}
  \begin{tabular}{ccccccccccccc}
    \hline
    Atomic Sp. & HD~4321 & HD~4321 & HD~125658 & HD~125658 & HD~137928 & HD~137928 & HD~154713 & HD~154713 & HD~159834 & HD~159834 & Solar\\
             & log $\epsilon$ & [x/H] & log $\epsilon$ & [x/H] & log $\epsilon$ & [x/H] & log $\epsilon$ & [x/H] & log $\epsilon$ & [x/H] & log $\epsilon_{\sun}$ \\
    \hline
    C~I      & 8.72 & 0.20$\pm$0.07 (11) & 8.72 & 0.20$\pm$0.09(7) & 8.73 & 0.21$\pm$0.08(2) & 8.42 & -0.10$\pm$0.11(3) & 8.46 & -0.06$\pm$0.08(11) & 8.52\\   
    O~I      & 8.88 & 0.05$\pm$0.25 (1) & 8.58 & -0.25$\pm$0.18(2) & 9.03 & 0.20$\pm$0.01(1) & $\dots$ & $\dots$ & 8.72 & -0.11$\pm$0.05(1) & 8.83\\
    Na~I     & 6.73 & 0.40$\pm$0.12 (3) & 6.71 & 0.38$\pm$0.10(2) & $\dots$& $\dots$ & $\dots$ & $\dots$ & 6.40 & 0.07$\pm$0.14(2) & 6.33\\
    Mg~I     & 7.57 & -0.01$\pm$0.15 (4) & 7.73 & 0.15$\pm$0.14(3) & 7.78 & 0.20$\pm$0.21(4) & 7.25 & -0.19$\pm$0.22(4) & 7.49 & -0.09$\pm$0.13(4) & 7.58\\
    Mg~II    & $\dots$ & $\dots$ & $\dots$ & $\dots$ & $\dots$ & $\dots$ & $\dots$ & $\dots$ & 7.62 & 0.04$\pm$0.10(2) & 7.58\\
    Al~II    & $\dots$ & $\dots$ & 6.70 &0.23$\pm$0.03(1) & $\dots$ & $\dots$ & $\dots$ & $\dots$ & $\dots$ & $\dots$ & 6.47\\
    Si I     & 7.63 & 0.08$\pm$0.14 (3) & 7.98 & 0.43$\pm$0.12(7) & $\dots$ & $\dots$ & $\dots$ & $\dots$ & 7.75 & 0.20$\pm$0.11(5) & 7.55\\
    Si~II    & 7.56 & 0.01$\pm$0.18 (1) & $\dots$ & $\dots$ & 7.77 & 0.22$\pm$0.17(3) & 7.43 & -0.12$\pm$0.17(3) & $\dots$ & $\dots$ & 7.55\\
    S~I      & 7.66 & 0.33$\pm$0.12 (5) & 7.71 & 0.38$\pm$0.11(2) & 7.69 & 0.36$\pm$0.10(1) & 7.64 & 0.31$\pm$0.10(1) & 7.41 & 0.08$\pm$0.09(1) & 7.33\\
    K~I       & $\dots$ & $\dots$ & $\dots$ & $\dots$ & $\dots$ & $\dots$ & $\dots$ & $\dots$ & 5.13 & 0.01$\pm$0.13(1) & 5.12\\
    Ca I     & 6.20 & -0.16$\pm$0.16 (10) & 6.25 & -0.11$\pm$0.15(11) & 6.47 & 0.11$\pm$0.16(3) & 6.05 & -0.31$\pm$0.18(6) & 6.28 & -0.08$\pm$0.17(12) & 6.36\\
    Ca~II    & 6.31 & -0.05$\pm$0.05 (2) & 6.27 & -0.09$\pm$0.04(2) & $\dots$ & $\dots$ & $\dots$ & $\dots$ & $\dots$ & $\dots$ & 6.36\\
    Sc II$^{HFS}$ & 2.62 & -0.55$\pm$0.14 (5) & 2.76 & -0.41$\pm$0.18(3) & 2.79 & -0.38$\pm$0.19(2) & 2.63 & -0.54$\pm$0.11(2) & 2.86 & -0.27$\pm$0.12(5) & 3.17\\
    Ti~II    & 5.23 & 0.21$\pm$0.22 (18) & 5.39 & 0.37$\pm$0.10(17) & 5.31 & 0.29$\pm$0.18(11) & 5.21 & 0.19$\pm$0.23(11) & 5.39 & 0.37$\pm$0.22(22) & 5.02\\
    V~II$^{HFS}$  & 4.18 & 0.18$\pm$0.12 (2) & 4.42 & 0.42$\pm$0.08(2) & 4.52 & 0.52$\pm$0.13(1) & 4.21 & 0.21$\pm$0.11(1) & 4.31 & 0.31$\pm$0.12(3) & 4.00\\
    Cr~I     & 5.84 & 0.17$\pm$0.15 (5) & 6.17 & 0.50$\pm$0.15(3) & $\dots$ & $\dots$ & 5.79 & 0.12$\pm$0.27(2) & $\dots$ & $\dots$ & 5.67\\
    Cr~II    & 5.87 & 0.20$\pm$0.11 (24) & 6.07 & 0.40$\pm$0.07(19) & 6.15 & 0.48$\pm$0.12(18) & 5.83 & 0.16$\pm$0.25(5) & 5.86 & 0.19$\pm$0.12(21) & 5.67\\
    Mn~I$^{HFS}$  & 5.62 & 0.23$\pm$0.18 (2) & 5.91 & 0.32$\pm$0.13(4) & 5.81 & 0.42$\pm$0.15(3) & 5.58 & 0.19$\pm$0.16(1) & 5.60 & 0.21$\pm$0.13(3) & 5.39\\
    Mn II$^{HFS}$ & 5.61 & 0.22$\pm$0.03 (1) & 5.98 & 0.39$\pm$0.08(2) & $\dots$ & $\dots$ & $\dots$ & $\dots$ & $\dots$ & $\dots$ & 5.39\\
    Fe~I     & 7.64 & 0.14$\pm$0.15 (106) &7.93 & 0.43$\pm$0.13(129) & 7.92 & 0.42$\pm$0.15(64) & 7.68 & 0.18$\pm$0.20(59) & 7.66 & 0.16$\pm$0.16(108) & 7.50\\
    Fe~II    & 7.62 & 0.12$\pm$0.15 (49) & 7.89 & 0.39$\pm$0.08(50) & 7.91 & 0.41$\pm$0.14(40) & 7.68 & 0.18$\pm$0.21(36) & 7.65 & 0.15$\pm$0.16(59) & 7.50\\
    Ni~I     & 6.46 & 0.21$\pm$0.14 (6) & 6.95 & 0.70$\pm$0.12(13) & 6.90 & 0.65$\pm$0.13(4) & 6.47 & 0.22$\pm$0.15(6) & 6.55 & 0.30$\pm$0.13(10) & 6.25\\
    Cu~I     & $\dots$ & $\dots$ & 4.93 & 0.72±0.13(1) & $\dots$ & $\dots$ & $\dots$ & $\dots$ & 4.53 & 0.32±0.21(2) & 4.21\\
    Zn~I     & 5.08 & 0.48$\pm$0.14 (2) & 5.31 & 0.71$\pm$0.12(2) & 5.38 & 0.78$\pm$0.13(1) & 4.79 & 0.19$\pm$0.14(1) & 5.05 & 0.45$\pm$0.17(2) & 4.60\\
    Sr~II    & 3.59 & 0.62$\pm$0.39 (2) & 3.46 & 0.49$\pm$0.47(1) & 3.67 & 0.70$\pm$0.51(1) & 3.82 & 0.85$\pm$0.48(1) & 3.94 & 0.97$\pm$0.51(1) & 2.97\\
    Y~II$^{HFS}$  & 2.56 &0.32$\pm$0.11 (2) & 3.51 & 1.07$\pm$0.10(5) & 3.26 & 1.02$\pm$0.15(3) &  $\dots$ &  $\dots$ & 2.81 & 0.57$\pm$0.17(7) & 2.24\\
    Zr~II    & 3.23 & 0.63$\pm$0.14 (2) & 4.08 & 1.48$\pm$0.13(3) & 3.81 & 1.21$\pm$0.11(3) & 3.28 & 0.68$\pm$0.18(2) & 3.51 & 0.91$\pm$0.10(3) & 2.60\\
    Ba~II$^{HFS}$ & 3.06 & 0.83$\pm$0.27 (3) & 3.61 & 1.18$\pm$0.23(3) & 3.36 & 1.23$\pm$0.22(2) & 2.92 & 0.79$\pm$0.29(3) & 3.02 & 0.89$\pm$0.30(2) & 2.13\\
    La~II    & 2.45 & 1.28$\pm$0.14 (1) & 2.58 & 1.41$\pm$0.12(1) & $\dots$ & $\dots$ & $\dots$ & $\dots$ & 2.68 & 1.51$\pm$0.19(1) & 1.17\\
    Ce~II    & 1.94 & 0.36$\pm$0.12 (1) & 2.99 & 1.41$\pm$0.11(1) & $\dots$ & $\dots$ & $\dots$ & $\dots$ & 2.77 & 1.19$\pm$0.15(4) & 1.58\\
    Nd~II    & 2.55 & 1.05$\pm$0.20 (2) & $\dots$ & $\dots$ & $\dots$ & $\dots$ & $\dots$ & $\dots$ & $\dots$ & $\dots$ & 1.50\\
    Sm~II    & 2.97 & 1.96$\pm$0.14 (1) & $\dots$ & $\dots$ & $\dots$ & $\dots$ & $\dots$ & $\dots$ & 2.59 & 1.58$\pm$0.13(1) & 1.01\\
    Eu~II    & $\dots$ & $\dots$ & $\dots$ & $\dots$ & $\dots$ & $\dots$ & $\dots$ & $\dots$ & 1.42 & 0.91$\pm$0.14(1) & 0.51\\
    Gd~II    & 2.90 & 1.78$\pm$0.10 (1) & $\dots$ & $\dots$ & $\dots$ & $\dots$ & $\dots$ & $\dots$ & $\dots$ & $\dots$ & 1.12\\
    \hline
  \end{tabular}
 \end{table}
\end{landscape}

\section{EVOLUTIONARY STATUS}
\label{sec:hr_dia}
To estimate the evolutionary status of the target stars, we precisely pinpointed their positions on the HR diagram. This is of particular importance as it allowed us to investigate the region occupied by the Am stars. With the help of isochrones and evolutionary tracks with solar metallicity \citep{Bressan2012}, we estimated the stellar masses and ages. The luminosity ($\text{L/L}_{\sun}$) was calculated using the standard technique. We obtained $\text{E(B-V)}$ from the work of \citet{Kervella2019} and computed extinction with the standard extinction law 
\begin{equation}
 [{A}_\text{V} \text{= 3.1 × E(B-V)}]
\end{equation}
The bolometric correction BC was calculated using the temperature-dependent function of \citet{Torres2010}. The absolute magnitudes (M$_V$) were determined using the Gaia parallaxes \citep{Gaia2020}. The errors in luminosities are from uncertainties in ${T}_\text{eff}$, V-band indices, and Gaia parallaxes. Using the Stefan–Boltzmann law as in \citet{Montambaux2018}, we calculated the radius from the ${T}_\text{eff}$ and log ($\text{L/L}_{\sun}$) values. The $\text{E(B-V)}$ values and calculated parameters are listed in Table~\ref{tab:hr_diagram}. The determined masses, ages and radii are listed in Table~\ref{tab:params}.

\begin{table*}
	\centering
	\caption{The logarithmic effective temperature log ${T}_\text{eff}$ (this study), $\text{E(B-V)}$ (\citet{Kervella2019}), bolometric correction BC (\citet{Torres2010}), absolute magnitude, M$_V$ (this study), parallax, $\pi$, \citep{Gaia2020} and luminosity, log($\text{L/L}_{\sun}$) (this study).}
	\label{tab:hr_diagram}
	\begin{tabular*}{\textwidth}{@{}l@{\hspace*{40pt}}c@{\hspace*{40pt}}c@{\hspace*{40pt}}c@{\hspace*{40pt}}c@{\hspace*{40pt}}c@{\hspace*{40pt}}c@{\hspace*{40pt}}c@{}}
		\hline
		Star name & log ${T}_\text{eff}$ &  $\pi$ (mas)& $\text{E(B-V)}$ & BC & M$_V$ & log($\text{L/L}_{\sun}$) \\
		\hline
		HD~2924 & 3.9370 [0.0075]   & 4.3978 [0.0292] & 0.040 & -0.0156 & -0.2078 & 1.9453 [0.0100]\\
		HD~4321 & 3.9294 [0.0076]   & 5.9672 [0.0209] & 0.034 & -0.0001 &  0.2834 & 1.7427 [0.0078]\\
		HD~26553 & 3.9217 [0.0077]] & 1.5391 [0.0370] & 0.170 &  0.0120 & -3.5107 & 3.2555 [0.0215]\\
		HD~125658 & 3.9445 [0.0072] &11.9629 [0.1441] & 0.007 & -0.0341 &  1.6918 & 1.1929 [0.0147]\\
		HD~137928 & 3.9590 [0.0071] & 8.1997 [0.0243] & 0.014 & -0.0341 &  0.9556 & 1.5051 [0.0111]\\
		HD~154713 & 3.9217 [0.0077] & 6.3445 [0.0227] & 0.021 &  0.0120 &  0.3878 & 1.6961 [0.0070]\\
		HD~159834 & 3.9217 [0.0077] & 6.4269 [0.1041] & 0.023 &  0.0120 &  0.0687 & 1.8237 [0.0153]\\
		\hline
	\end{tabular*}
\end{table*}

\begin{figure}
	\includegraphics[width=\columnwidth]{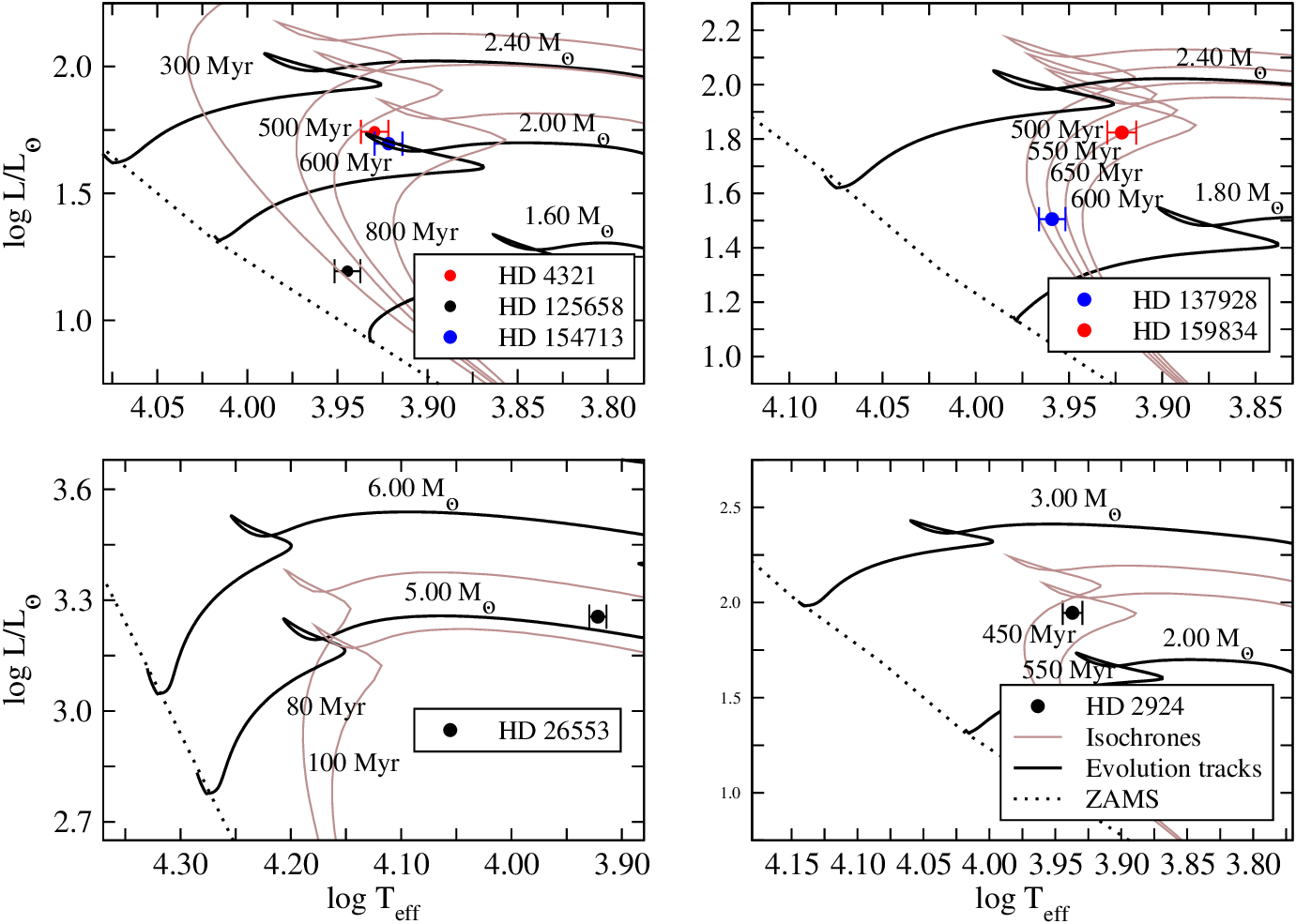}
    \caption{The position of target stars on the HR diagram. Evolutionary tracks and isochrones are from \citet{Bressan2012}.}
    \label{fig:hrdia}
\end{figure}

\begin{table}
	\centering
	\caption{The masses, ages, and radii of the target stars.}
	\label{tab:params}
	\begin{tabular}{lccr} 
		\hline
		Star name & Mass [$M_{\sun}$] & Age [Myr]& Radius [$R_{\sun}$] \\
	
		\hline
		HD~2924 & 2.40 [0.09] & 490 [40]  & 04.20 [0.40] \\
		HD~4321 & 2.00 [0.08] & 625 [50]  & 03.40 [0.30] \\
		HD~26553 & 5.10 [0.30] & 090 [05] & 20.30 [2.20]\\
		HD~125658 & 1.75 [0.06] & 370 [40] & 01.80 [0.20]\\
		HD~137928 & 2.00 [0.07] & 550 [50] & 02.30 [0.25]\\
		HD~154713 & 2.00 [0.08] & 675 [60] & 03.40 [0.30]\\
		HD~159834 & 2.20 [0.07] & 600 [50] & 03.90 [0.40]\\
		\hline
	\end{tabular}
\end{table}

\section{Results}
\label{sec:res}
\subsection{Normal A-type stars}  
\subsubsection{HD~2924}
This star was classified as spectral type A2IV \citep{Cowley1969}. \citet{Zorec2012} measured ${T}_\text{eff}$ as 8730\,K (log ${T}_\text{eff}$ = 3.941) and reported log ($\text{L/L}_{\sun}$) = 2.3083, mass = 3.06 $M_{\sun}$, and a rotational velocity of 30~km\,s$^{-1}$. \citet{McDonald2012} measured the ${T}_\text{eff}$ as 7800\,K and reported a luminosity of 134.59 (log($\text{L/L}_{\sun}$) = 2.129). \citet{Duflot1995} measured a radial velocity of 1.9 km\,s$^{-1}$, whereas \citet{Bruijne2012} and \citet{Brandt2021} reported a radial velocity of 2.1~km\,s$^{-1}$. 
In this study, we measured the fundamental parameters of HD~2924 as ${T}_\text{eff}$ = 8650\,K, log $g$ = 3.7 and $\xi$ = 2.60 km\,s$^{-1}$. We derived $v$ sin $i$ and ${V}_\text{helio}$ as 30~km\,s$^{-1}$ and 1.90~km\,s$^{-1}$, respectively. The abundance of calcium is 0.13\,dex and scandium -0.08\,dex. The iron-peak and heavy elements do not show significant overabundance as we expect for Am stars, hence we classified HD~2924 as a normal A-type star. The elemental abundance distribution of HD~2924 is shown in Fig.~\ref{fig:normal}. We estimated the mass, age, and radius as 2.40$\pm$0.09~$M_{\sun}$, 490$\pm$60~Myr, and 4.20$\pm$0.40~$R_{\sun}$, respectively. The difference between the calculated luminosity in this study and the luminosity given in \citet{Zorec2012} is largely due to the parallax values used in the calculation of luminosities.
 
\subsubsection{HD~26553}
HD 26553 was listed as A4III by \citet{Cowley1969}. \citet{Bruijne2012} and \citet{Brandt2021} reported a radial velocity of -24.9~km\,s$^{-1}$. \citet{Royer2002} measured $v$ sin $i$ as 18~km\,s$^{-1}$. We determined the atmospheric parameters of HD~26553 as follows; ${T}_\text{eff}$ = 8350~K, log $g$ = 2.2, and $\xi$ = 1.9~km\,s$^{-1}$. We measured $v$ sin $i$ and ${V}_\text{helio}$ as 12~km\,s$^{-1}$ and 22~km\,s$^{-1}$, respectively. With [Ca/H] = 0.04~dex, [Sc/H] = -0.13~dex and an average abundance of iron-peak and heavy elements, the star exhibits solar-like abundances. The chemical abundance pattern, shown in Fig.~\ref{fig:normal}, does not show Am characteristics and is consistent with the pattern of normal A-type stars. The physical parameters were estimated as mass = 5.1±0.3~$M_{\sun}$, age = 90$\pm$5~Myr and radius = 20.3±2.2~$R_{\sun}$. The HR diagram reveals that HD~26553 is a post-TAMS star.

\subsection{Classical Am stars}
\subsubsection{HD~4321}
HD~4321 was listed as an A2III star in the HD catalogue of \citet{Cannon1993}. The star’s radial velocity was measured as -8.3~km\,s$^{-1}$ \citep{Bruijne2012, Brandt2021}. \citet{McDonald2012} reported an effective temperature of 7787~K and a luminosity of 50.05 $L_{\sun}$ (log ($\text{L/L}_{\sun}$)=1.70). \citet{Royer2002} measured $v$ sin $i$ as 24~km\,s$^{-1}$. We measured the star’s ${T}_\text{eff}$, log $g$, and $\xi$ as 8500~K, 4.00, and 3.00 ~km\,s$^{-1}$, respectively. We measured the $v$ sin $i$ and ${V}_\text{helio}$ of HD~4321 as 24~km\,s$^{-1}$ and -7.6~km\,s$^{-1}$. Its [Ca/H] and [Sc/H] values were found to be underabundant with respect to the solar abundances by –0.16 and –0.55 dex. The star exhibits an overabundance (on average) in iron-peak and rare-earth elements with respect to the solar abundances. Its chemical abundance pattern, shown in Fig.~\ref{fig:classic}, implies a classical Am star. The mass of HD~4321 was estimated as 2.00$\pm$0.08~$M_{\sun}$. We estimated the age and the radius as 625$\pm$60~Myr and 3.40$\pm$0.30~$R_{\sun}$, respectively.

\subsubsection{HD~125658}
This target was first noted as an Am-type candidate by \citet{Bertaud1958}. \citet{Cowley1969} classified HD~125658 as A5III. It was listed as an A2 star in the HD catalogue \citep{Cannon1993}. \citet{Bruijne2012} and \citet{Brandt2021} measured the radial velocity as 1.3~km\,s$^{-1}$. \citet{McDonald2012} derived the ${T}_\text{eff}$ as 8123~K and the luminosity as 17.9~$L_{\sun}$ (log ($\text{L/L}_{\sun}$) = 1.25). \citet{David2015} reported ${T}_\text{eff}$ as 8430~K, log $g$ as 4.27, the mean age from 1D posterior as 581~Myr, and the mass as 1.9~$M_{\sun}$. From 2D linear interpolation, the age and mass were measured as 363~Myr and 1.78~$M_{\sun}$, respectively. \citet{Royer2002} measured a $v$ sin $i$ of 26~km\,s$^{-1}$. We measured the atmospheric parameters of HD~125658 as ${T}_\text{eff}$ = 8800~K, log $g$ = 4.60, and $\xi$ = 5.07~km\,s$^{-1}$. Its $v$ sin $i$ and heliocentric radial velocity ${V}_\text{helio}$ were derived as 26~km\,s$^{-1}$ and 1.66~km\,s$^{-1}$. HD~125658 has characteristic abundances consistent with the abundance of classical Am-type stars. The underabundance of Ca at -0.11 and Sc at -0.41 and the overabundance pattern in iron-peak and heavy elements with respect to the solar abundances are shown in Fig.~\ref{fig:classic}. We note that the star has significantly higher iron, nickel, copper, and zinc abundances than the other two Am-type samples. The mass, age, and radius were estimated as 1.75$\pm$0.06~$M_{\sun}$, 370$\pm$40~Myr, and 1.80~$\pm$0.20$~R_{\sun}$, respectively. 

\subsubsection{HD~154713}
HD~154713 was listed as an A0 star in the HD catalogue \citep{Cannon1993}. The spectral type of HD~154713 was classified as A3IV \citep{Cowley1969}. \citet{Zorec2012} measured ${T}_\text{eff}$ as 8550~K (log ${T}_\text{eff}$ = 3.932) and reported log ($\text{L/L}_{\sun}$) = 1.81, mass = 2.52~$M_{\sun}$, and a rotational velocity of 39~km\,s$^{-1}$. \citet{McDonald2012} measured the ${T}_\text{eff}$ as 7954~K and reported a luminosity of 47.48~$L_{\sun}$ (log ($\text{L/L}_{\sun}$) = 1.679). \citet{Bruijne2012} and \citet{Brandt2021} reported a radial velocity of -8.7~km\,s$^{-1}$. We determined the atmospheric parameters of HD~154713 as ${T}_\text{eff}$ = 8350~K, log $g$ = 3.50, and $\xi$ = 3.00~km\,s$^{-1}$. We derived the values of $v$ sin $i$ and ${V}_\text{helio}$ as 39~km\,s$^{-1}$ and -6.30~km\,s$^{-1}$, respectively. The abundance of calcium is -0.31~dex and scandium -0.54~dex. The iron-peak and heavy element abundances are slightly overabundant with respect to solar; hence, HD~154713 is consistent with classical Am-type stars. The abundance pattern is shown in Fig.~\ref{fig:classic}. We estimated the mass, age, and radius as 2.00$\pm$0.08~$M_{\sun}$, 675$\pm$60~Myr, and 3.40$\pm$0.40~$R_{\sun}$, respectively.

\subsection{Marginal Am stars}

\subsubsection{HD~137928}
This star was ascribed a spectral type A2 IV \citep{Cowley1969}. \citet{Zorec2012} measured ${T}_\text{eff}$ as 8649 K (log ${T}_\text{eff}$ = 3.937) and reported log ($\text{L/L}_{\sun}$) = 1.744, mass = 2.45~$M_{\sun}$, and $v$ sin $i$ = 34~km\,s$^{-1}$. \citet{McDonald2012} measured the ${T}_\text{eff}$ as 8637~K and reported a luminosity of 32.62~$L_{\sun}$ (log ($\text{L/L}_{\sun}$) = 1.513). \citet{Bruijne2012} and \citet{Brandt2021} reported a radial velocity of -4.90km\,s$^{-1}$. \citet{Kervella2019} measured the radial velocity as -4.985km\,s$^{-1}$. We determined the ${T}_\text{eff}$ of HD~137928 as 9100~K. The log $g$ and $\xi$ were measured as 3.8 and 4.03~km\,s$^{-1}$, respectively. The $v$ sin $i$ was determined as 34~km\,s$^{-1}$ and ${V}_\text{helio}$ as -3.9~km\,s$^{-1}$. The abundance of calcium is 0.11~dex and that of scandium is -0.38~dex. The Ca abundance is nearly solar-like, while Sc is slightly underabundant. Iron-peak and heavy elements have an overabundance implying marginal Am star abundance characteristics. HD~137928 has high iron, nickel, and zinc abundances. This makes it a metal-rich star. The elemental abundance pattern of HD~137928 is shown in Fig.~\ref{fig:marginal}. Its mass was determined as 2.00$\pm$0.07~$M_{\sun}$. HD~137928’s radius was measured as 2.30$\pm$0.25~$R_{\sun}$ at the age of 550$\pm$50~Myr. 

\subsubsection{HD~159834}
This star was first classified as A7III by \citet{Eggen1962}. \citet{Cowley1969} reported the spectral type as A7IV. HD~159834 was listed as an A2 star in the HD catalogue \citep{Cannon1993}. \citet{Zorec2012} measured its ${T}_\text{eff}$ as 7925~K (log ${T}_\text{eff}$ = 3.899) and reported a luminosity of log ($\text{L/L}_{\sun}$) = 1.6256, a mass of 2.29~$M_{\sun}$, and a rotational velocity of~km\,s$^{-1}$. \citet{McDonald2012} measured the ${T}_\text{eff}$ as 7724 K and reported a luminosity of 34.94~$L_{\sun}$ (log ($\text{L/L}_{\sun}$) = 1.54). \citet{Bruijne2012} and \citet{Brandt2021} reported a radial velocity of -17~km\,s$^{-1}$. \citet{Burkhart1991} measured the ${T}_\text{eff}$, log $g$ and [Fe/H] as 8030~K, 3.7, and 0.1~dex, respectively. \citet{Kervella2019} measured a radial velocity of –16.82~km\,s$^{-1}$. Finally, the atmospheric parameters of HD~159834 were determined as ${T}_\text{eff}$ = 8350~K, log $g$ = 3.9, and $\xi$ = 3.2~km\,s$^{-1}$. We measured $v$ sin $i$ and ${V}_\text{helio}$ as 14~km\,s$^{-1}$ and -16.20~km\,s$^{-1}$. The calcium abundance is –0.08~dex, while that of scandium is –0.27~dex. The abundances are close to solar, with scandium slightly underabundant. The iron-peak and heavy elements are moderately overabundant, implying a marginal Am star. The elemental abundance pattern is shown in Fig.~\ref{fig:marginal}. We determined the mass, age, and radius as 2.20$\pm$0.07~$M_{\sun}$, 600$\pm$50~Myr, and 3.90$\pm$0.4~$R_{\sun}$, respectively.

\section{Conclusion}
\label{sec:conc}
In this paper, we present a chemical abundance analysis of seven A-type stars, for which detailed abundance analyses are absent in the literature, to search for their chemical anomalies. These stars are slow-rotating A-type stars. Slow rotators were targeted because rotational velocity is linked to the mixing process of stellar materials that suppress the effect of chemical separation.
We first obtained fundamental atmospheric parameters and velocities for each target, including the effective temperature (${T}_\text{eff}$), surface gravity (log $g$), metallicity [Fe/H], microturbulent velocity ($\xi$), rotational velocity ($v$ sin $i$), and radial velocity (${V}_\text{helio}$). We then performed a detailed abundance analysis for each star, obtained their luminosity log ($\text{L/L}_{\sun}$), and pinpointed their position in the HR diagram. Using evolutionary tracks and isochrones, we estimated the masses and the ages of the targets. Using the Teff and luminosity values, we calculated the radii of the stars.
The present work is the first detailed abundance analysis of these targets. Six of the targets are main–sequence stars with some of them near to the Terminal-Age Main Sequence, while HD~26553 was revealed to be a post-TAMS giant. Based on the spectral analysis, the resulting chemical abundance pattern of five targets revealed that they are Am-type stars, with HD~4321, HD~125658, and HD~154713 classified as classical Am stars, while HD~137928 and HD~159834 were classified as marginal Am stars. The abundance patterns of HD~2924 and HD~26553 are consistent with those of normal A-type stars.
Two of our samples, HD~125658 (classified as a classical Am star) and HD~137928 (classified as a marginal Am star), have high iron, nickel, and zinc abundances. Their carbon, sodium, magnesium, sulphur, scandium, vanadium, manganese, yttrium, and barium abundances are very similar. They also have very similar atmospheric and physical parameters. Considering all these similarities, these two stars are among the most metal–rich Am stars. The HR diagram shows that Am stars populate the main sequence. 
On average based on the chemical separation theory, the main-sequence slow rotator HD~2924 is expected to exhibit Am characteristics. The presence of its non-Am characteristics poses a challenge on rotation effect to the hypothesis chemical separation, which shows that there is a need to study more of these targets to explore the origin of their peculiarity and their interior processes.

\section*{Acknowledgements}
The authors acknowledge the TÜBİTAK National Observatory (TUG) with the project ID 14BRTT150–671.
\section*{Data Availability}
The data underlying this article are available in the article and in its online supplementary material. Upon request, further data underlying this article will be shared on reasonable request to the corresponding author.


\bibliographystyle{mnras}
\bibliography{references} 



\bsp	
\label{lastpage}
\end{document}